\documentclass[a4paper,UKenglish,cleveref,autoref,thm-restate]{lipics-v2021}

\usepackage{tikz}
\usetikzlibrary{positioning}
\usetikzlibrary{arrows.meta}
\usetikzlibrary{shapes}
\usetikzlibrary{shapes.geometric}

\usepackage{tabularx}
\usepackage{multirow}
\usepackage[utf8]{inputenc}

\title{The Influence of Fraudulent AI-Generated Responses on Software Engineering Surveys} 

\Copyright{Ronnie de Souza Santos and Italo Santos and Maria Teresa Baldassarre and Cleyton Magalhães and Mairieli Wessel}

\author{Ronnie de Souza Santos}
{University of Calgary, Canada}
{ronnie.desouzasantos@ucalgary.ca}
{https://orcid.org/0000-0002-2631-4604}
{}

\author{Italo Santos}
{University of Hawaii at Manoa, USA}
{isantos3@hawaii.edu}
{https://orcid.org/0000-0002-7545-6104}
{}

\author{Maria Teresa Baldassarre}
{University of Bari, Italy}
{mariateresa.baldassarre@uniba.it}
{https://orcid.org/0000-0001-8589-2850}
{}

\author{Cleyton Magalhães}
{Universidade Federal Rural de Pernambuco (UFRPE), Brazil}
{cleyton.vanut@ufrpe.br}
{https://orcid.org/0009-0005-3051-7232}
{}

\author{Mairieli Wessel}
{Radboud University, Netherlands}
{mairieli.wessel@ru.nl}
{https://orcid.org/0000-0001-8619-726X}
{}

\authorrunning{de Souza Santos et al.}



\EventEditors{ESEM Editors}
\EventNoEds{1}
\EventLongTitle{20th International Symposium on Empirical Software Engineering and Measurement}
\EventShortTitle{ESEM 2026}
\EventAcronym{ESEM}
\EventYear{2026}
\EventDate{October 4--9, 2026}
\EventLocation{Munich, Germany}
\EventLogo{}
\SeriesVolume{}
\ArticleNo{23}

\begin{document}

\maketitle

\begin{abstract}
\textbf{Background:} Large Language Models (LLMs) introduce new concerns regarding fraudulent or AI-assisted participation in software engineering surveys. \textbf{Aims:} This study investigates how suspicious or potentially AI-assisted responses may affect the validity of software engineering survey findings. \textbf{Method:} We conducted a secondary analysis of four software engineering survey datasets using manual identification of suspicious responses, automated AI-generated text detection, descriptive statistical analysis, and thematic analysis. We compared findings obtained from the original and manually cleaned datasets. \textbf{Results:} Quantitative findings generally remained stable after filtering suspicious responses, although some demographic and analytical variables showed moderate variation, affecting the interpretation of specific participant groups and contextual characteristics. In contrast, qualitative findings were more strongly influenced by changes in contextual framing, code prominence, and the nature of the evidence supporting interpretation, shaping how participants' experiences and study contexts were interpreted and characterized. \textbf{Conclusions:} AI-assisted participation may influence software engineering survey findings differently depending on the type of analysis being conducted. The findings reinforce the importance of combining multiple validation procedures, particularly in studies relying on open-ended responses.
\end{abstract}


\vspace{-5px}
\section{Introduction}
\label{sec:introduction}
\vspace{-5px}

Surveys are widely used in empirical software engineering to investigate developer practices, perceptions, and experiences across diverse organizational and educational contexts~\cite{pfleeger2001principles, kitchenham2002principles, punter2003conducting, ciolkowski2003practical, molleri2016survey, molleri2020empirically}. The increasing use of online recruitment platforms has expanded the reach of survey-based studies while also introducing concerns related to response authenticity, participant validity, and data quality~\cite{danilova2021you, lawlor2021suspicious, pratt2021strategies, johnson2024addressing, de2025investigation}. Previous research has shown that online surveys are susceptible to inattentive participation, fabricated responses, bots, and coordinated submissions capable of affecting empirical findings~\cite{lawlor2021suspicious, pratt2021strategies, zhang2022beyond, bell2023fraud}. Consequently, the literature recommends combining validation procedures such as attention checks, consistency verification, metadata inspection, and participant screening mechanisms to improve survey reliability~\cite{danilova2021you, lawlor2021suspicious, pratt2021strategies, zhang2022beyond, bonnamy2025survey}.

Recent advances in Large Language Models (LLMs) introduce additional methodological concerns for empirical software engineering research. LLMs are increasingly used in software engineering activities and research workflows, including code generation, documentation, literature reviews, and scientific writing~\cite{fan2023large, sallou2024breaking, lim2025llm, rane2023contribution, liang2024mapping, lecca2025applications, felizardo2024chatgpt}. Emerging studies also indicate that these systems can generate responses resembling human narratives and simulate participant behavior in survey contexts~\cite{steinmacher2024can, kaiser2025simulating}. As a result, AI-generated or substantially modified responses may no longer represent authentic participant experiences or perceptions~\cite{pinzon2024ai, de2025investigation}.

Although interest in LLMs in software engineering has grown, little research has examined their impact on the authenticity and validity of survey responses. Existing studies predominantly focus on researchers' use of LLMs or traditional forms of survey fraud~\cite{rane2023contribution, liang2024mapping, lecca2025applications, bell2023fraud, zhang2022beyond}. Therefore, we investigate the following research question (RQ): \textit{How can the use of LLMs by participants affect the authenticity and validity of data collected in software engineering surveys?} To answer this RQ, we analyze four software engineering survey datasets using a multi-stage detection protocol combining manual inspection, narrative characterization, and automated verification techniques. Our contributions include: (1) a methodological investigation of suspicious responses across different software engineering survey contexts, (2) a comparative analysis evaluating how suspicious responses influence both quantitative and qualitative findings across original and filtered datasets, and (3) methodological evidence regarding the limitations of automated AI detection tools and the continued importance of manual inspection in survey validation procedures.


\vspace{-5px}
\section{Background}
\label{sec:background}
\vspace{-5px}


\vspace{-5px}
\textbf{Survey Research in Software Engineering.} Surveys are widely used in empirical software engineering to collect information about developer practices, perceptions, and experiences that are difficult to observe in controlled settings~\cite{kitchenham2002principles, pfleeger2001principles, punter2003conducting}. 
Software engineering literature also provides methodological guidance on designing survey instruments, selecting appropriate participant populations, and reporting survey results in a transparent and systematic manner~\cite{ciolkowski2003practical, punter2003conducting}. Recruitment practices have evolved with the increasing use of online platforms and digital communication channels. Surveys are commonly distributed through professional mailing lists, social networks, developer communities, and specialized recruitment platforms~\cite{danilova2021you, bonnamy2025survey}. Platforms such as Prolific and similar participant marketplaces have been adopted in empirical studies because they provide access to large pools of respondents while enabling demographic and professional filtering during recruitment~\cite{bonnamy2025survey, douglas2023data, reid2022software, russo2022recruiting}. Despite these advantages, determining whether respondents belong to the intended population of software practitioners remains a methodological challenge~\cite{wagner2020challenges, reid2022software, russo2022recruiting}. The literature therefore recommends the use of screening questions and experience-related measures to assess respondents' professional background and expertise~\cite{ciolkowski2003practical, punter2003conducting}. In addition, sampling strategies and recruitment approaches may influence the representativeness of survey samples, particularly when participation is voluntary or mediated through online platforms~\cite{baltes2022sampling, wagner2020challenges, reid2022software, russo2022recruiting}. These characteristics of online survey recruitment also introduce risks related to response authenticity and data quality. 
Consequently, concerns about fraudulent participation, inattentive responses, and other forms of data contamination have become increasingly relevant in survey-based empirical research~\cite{lawlor2021suspicious, pratt2021strategies, bell2023fraud, zhang2022beyond}.

\textbf{Fraud and Data Quality in Online Surveys.} Ensuring data validity is an important concern in online survey research as invalid or fraudulent participation can compromise the reliability of empirical findings~\cite{lawlor2021suspicious, johnson2024addressing}. Online surveys are susceptible to inattentive responses, repeated participation attempts, fabricated answers, and automated submissions that do not represent the genuine experiences of the intended participant population~\cite{lawlor2021suspicious, pratt2021strategies, bell2023fraud}. Fraudulent or low-quality responses can distort empirical findings by introducing noise or systematic bias into survey datasets~\cite{zhang2022beyond}. Automated submissions generated by bots or coordinated responses from survey farms may artificially inflate sample sizes while reducing the reliability of collected data~\cite{bell2023fraud, zhang2022beyond}. These issues are important in online survey research because invalid participation may affect both quantitative analyses and qualitative interpretations, potentially influencing conclusions derived from the data. To address these risks, prior literature proposes several validation and fraud-detection mechanisms to identify suspicious participation patterns~\cite{lawlor2021suspicious, pratt2021strategies}. Common approaches include consistency checks across related questions, attention-verification items, completion-time analysis, screening questions related to domain expertise, and the use of technical metadata to identify repeated participation attempts~\cite{lawlor2021suspicious, pratt2021strategies, zhang2022beyond}. Because fraudulent responses may resemble legitimate participation, the literature commonly recommends combining multiple validation strategies to improve the reliability and quality of survey datasets~\cite{zhang2022beyond, johnson2024addressing}.

\textbf{LLMs in Empirical Software Engineering Research.} While existing validation mechanisms are designed to detect inattentive participation, automated submissions, and other traditional forms of survey fraud, recent advances in AI introduce additional methodological considerations for empirical research. LLMs are capable of generating coherent and plausible responses across diverse topics~\cite{fan2023large, sallou2024breaking, lim2025llm}, and are increasingly being used to support research activities such as literature reviews, data synthesis, and scientific writing~\cite{rane2023contribution, liang2024mapping, lecca2025applications, felizardo2024chatgpt}. Emerging studies also indicate that LLMs can generate responses that resemble narratives typically produced by human participants in surveys and experiments~\cite{steinmacher2024can, kaiser2025simulating}. These capabilities raise concerns for studies that rely on self-reported data, since AI-generated or substantially modified responses may not reflect authentic experiences, perceptions, or practices~\cite{pinzon2024ai}. Recent investigations have documented cases in which AI tools were used to fabricate or modify survey responses submitted during online data collection~\cite{de2025investigation}, suggesting that AI tools may be used to fabricate survey data while appearing to reflect real participation. Despite increasing attention to researchers' use of LLMs, little attention has been paid to how participants' use of AI tools may influence the authenticity and reliability of survey data in empirical software engineering research.

\vspace{-5px}
\section{Method}
\label{sec:method}
\vspace{-5px}


We adopted a meta-scientific design to investigate a methodological issue in empirical software engineering research through the secondary analysis of existing survey datasets. Meta-science examines research practices, methodological procedures, and validity threats rather than focusing directly on software engineering phenomena~\cite{ralph2020empirical}. Specifically, we analyze the authenticity of survey responses and assess whether potentially LLM-generated responses influence empirical findings. To achieve this, we reanalyzed four previously collected survey datasets using secondary data analysis, a method that enables researchers to revisit existing data to address new research questions or methodological concerns~\cite{cheng2014secondary, johnston2014secondary, ruggiano2019conducting}.



\vspace{-5px}
\subsection{Study Contexts and Survey Datasets}
\vspace{-5px}

We conducted a secondary analysis using datasets from four questionnaire surveys we previously conducted in different software engineering contexts between 2025 and 2026. In these surveys, we investigated topics related to software engineering education and practice using both open- and closed-ended questions administered via online survey instruments. Although the surveys addressed distinct RQs, they share methodological characteristics typical of empirical software engineering surveys. Each survey was designed to collect participant perceptions, experiences, or practices through a combination of qualitative responses and structured questionnaire items, enabling comparative analysis of response patterns across datasets. The surveys investigated the following topics:

\begin{itemize}
\item \textbf{Technology education in the Global South:} a survey investigating challenges, opportunities, and educational practices related to technology and computing education in Global South contexts (16 questions).
\item \textbf{Student cheating using AI tools:} a survey investigating how students use generative AI systems in academic contexts and how such tools influence academic integrity practices (21 questions).
\item \textbf{Safe spaces in software engineering education:} a survey investigating how students define and perceive safe spaces in academic environments and how inclusive practices influence learning experiences (12 questions).
\item \textbf{LLM usage among software developers:} a survey investigating perceptions and experiences related to reliance on LLMs during software development activities (18 questions).
\end{itemize}

These four datasets provide diverse survey contexts while maintaining comparable methodological structures, including combinations of open-ended and closed-ended questions related to software engineering education and practice. This consistency allowed comparative analysis across different survey topics while preserving similar data collection and validation procedures between datasets. 

In all four surveys, our primary recruitment strategy involved the Prolific platform. Prolific is commonly used in empirical research to recruit participants for online studies and provides several mechanisms to support data quality \cite{reid2022software, russo2022recruiting}, including participant approval rate thresholds, demographic filters, and attention checks embedded in surveys. In each survey, we configured participant eligibility criteria and implemented multiple validation procedures intended to reduce inattentive participation, automated submissions, and low-quality responses. These procedures included participant approval rate thresholds available through Prolific, demographic and background screening criteria related to software engineering or computing experience, consistency verification across demographic questions, and attention validation questions embedded throughout the surveys \cite{reid2022software, russo2022recruiting, proli}. We also included domain-specific verification questions related to programming and software engineering knowledge to identify participants without the expected technical background for the studies. In addition, we manually inspected responses to identify indicators commonly associated with inattentive participation, including contradictory demographic information, duplicated narrative structures, repeated text patterns, nonsensical answers, and unusually generic open-ended responses.

Despite the use of these platform-level quality controls and validation procedures, we identified a number of responses demonstrating characteristics consistent with AI-generated text. This observation motivated the secondary analysis conducted in this study. More specifically, we analyzed the datasets to estimate the prevalence of suspicious responses and assess whether the presence of such responses influences empirical findings derived from survey-based studies. To preserve the double blind review process, we do not provide citations to the original survey studies in this manuscript. Some of these studies are currently under review, while others are already published but could still compromise reviewer anonymity if cited during the evaluation process. The corresponding references will be included in the final camera-ready version of the paper after the review process concludes.

\vspace{-5px}
\subsection{Data Selection}
\vspace{-5px}

To support a systematic comparison between the original and cleaned datasets, we selected a subset of analytical questions and demographic variables from each survey. The selection strategy was designed to include both qualitative and quantitative data commonly used in empirical software engineering research. Specifically, we selected demographic variables used to characterize participant populations, closed-ended questions representing central analytical constructs, and the main open-ended questions, also related to the central analytical construct in the research, that required reflective responses. Two authors participated in this process, and conflicts were resolved in agreement meetings, which could include a third author. 

\begin{itemize}

\item \textbf{Demographic variables.}  
We selected the demographic or background variables from each survey to evaluate whether the presence of suspicious responses affected the demographic composition of the survey sample. Demographic variables describe characteristics of participants and are commonly used to contextualize survey samples in empirical software engineering research. In the survey on technology education in the Global South, we analyzed gender, ethnicity, sexual orientation, and year of study. In the survey investigating inappropriate LLM use in academic contexts, we analyzed gender, ethnicity, and year of study. In the safe spaces survey, we analyzed gender, ethnicity, and year of study. In the survey on LLM reliance among software developers, we analyzed gender and years of professional experience.

\item \textbf{Key closed-ended analytical questions.}  
We also selected closed-ended questions from each survey that represented central constructs investigated in the original studies. These questions were designed as multiple-choice items with predefined response options. Because participants select among fixed alternatives, these questions produce structured data that can be analyzed quantitatively and allow comparison of response distributions across datasets. In the survey on technology education in the Global South, we analyzed questions related to financial support and first-generation university attendance status. In the survey investigating inappropriate LLM use in academic contexts, we analyzed questions related to training on LLMs, emotional responses associated with LLM use, and factors contributing to the use of LLMs in ways that were not permitted in coursework. In the safe spaces survey, we analyzed questions related to awareness of safe spaces and perceptions of exclusion or insecurity in software engineering education. In the survey on LLM reliance among software developers, we analyzed questions related to frequency of thinking about LLMs during work activities, comfort using generated code without human oversight, and formal LLM training background.

\item \textbf{Core open-ended analytical questions.}  
Finally, we selected one open-ended question from each survey that required participants to describe experiences, motivations, or challenges related to the topic of the study. These questions typically require reflective explanations and personal narratives. Such questions are particularly relevant in the context of this study because suspicious responses were most frequently observed in narrative answers requiring elaborated textual explanations \cite{de2025investigation}. These responses are also commonly used as the basis for qualitative analysis in empirical software engineering studies. The selected questions correspond to the central constructs investigated in each survey. In the survey on technology education in the Global South, we selected the question asking participants to describe a problem or challenge they experienced while pursuing technology education. In the survey investigating inappropriate LLM use in academic contexts, we selected the question asking participants to describe a situation in their software engineering program where they used an LLM in a way that was not aligned with course rules or expectations. In the safe spaces survey, we selected the question asking participants to describe how they understand or experience a safe space in software engineering education. In the survey investigating LLM reliance among software developers, we selected the question asking participants to describe a specific situation in which they used a LLM to assist with a programming task.

\end{itemize}

Across the four surveys, this selection strategy resulted in the analysis of 24 questions. This scope provides sufficient analytical coverage to evaluate the influence of suspicious responses while maintaining a tractable and systematic comparison across datasets.

\vspace{-5px}
\subsection{Data Analysis}
\vspace{-5px}

After selecting the questions for comparative analysis, we conducted a multi-stage qualitative and quantitative analysis procedure to evaluate how suspicious responses influence survey findings. The analysis process consisted of the following steps:

\begin{itemize}

\item \textbf{Manual identification of suspicious responses.}  We first analyzed the original datasets and manually identified suspicious responses through independent classification performed by two researchers. The researchers classified responses as suspicious or non-suspicious and subsequently discussed disagreements until consensus was achieved. The manual identification process was informed by patterns reported in prior exploratory work on AI-generated responses in software engineering surveys~\cite{de2025investigation}. During the review process, the researchers evaluated multiple indicators jointly rather than relying on a single characteristic in isolation. These indicators included repetitive narrative structures across participants, highly similar stylistic composition and rhetorical organization, reused or formulaic expressions, superficial personalization, duplicated semantic structures describing different situations with nearly identical discursive flow, unusually polished but homogeneous writing patterns, and inconsistencies or implausible contextual details within participant responses. Classification decisions emphasized recurring cross-response similarity patterns rather than isolated linguistic characteristics, and responses were not classified as suspicious solely because of writing quality or grammatical correctness.

\item \textbf{Automated AI-generated text detection.} We evaluated the same responses using the Scribbr AI Detector to estimate the probability that individual answers were AI-generated. During this process, the exact open-ended responses identified during manual analysis were copied and pasted into the detection tool without modification. The resulting probability estimates were recorded and compared with the manual classifications. Because AI detection tools provide probabilistic estimates rather than definitive classifications, these results were used to compare automated and manual identification procedures rather than as a definitive ground truth.

\item \textbf{Quantitative analysis of demographic and closed-ended questions.} We conducted descriptive statistical analyses \cite{george2018descriptive} using the demographic variables and selected closed-ended questions. For each variable and question, we calculated frequencies, percentages, and proportional distributions across response categories using the original datasets. We then repeated the same calculations using the manually cleaned datasets generated after removing suspicious responses. Comparing these distributions allowed identification of whether suspicious responses influenced the demographic characterization of participants or altered the distribution of quantitative survey findings. The demographic analyses included variables such as gender, ethnicity, year of study, sexual orientation, and professional experience, depending on the survey context. The closed-ended analyses included questions related to financial support, first-generation university attendance status, awareness of safe spaces, perceptions of exclusion or insecurity, training on LLMs, emotional responses associated with LLM use, comfort using generated code without human oversight, and frequency of thinking about LLMs during work activities.

\item \textbf{Qualitative analysis of open-ended responses.}   For the selected open-ended questions, we conducted thematic analysis following established qualitative research practices in empirical software engineering \cite{seaman1999qualitative,lenberg2024qualitative,clarke2017thematic}. First, two authors independently read the full set of responses for each selected question to become familiar with the data and identify recurring ideas. During this stage, the authors extracted quotes representing meaningful statements expressed by participants. These quotes served as the basic analytical units. Second, we assigned initial codes to each extracted quote. Codes were generated inductively from the data rather than from predefined categories. Third, we grouped related codes into broader conceptual categories representing recurring concepts across responses. Finally, we organized these categories into themes capturing the main patterns identified in the datasets. The two authors compared their coding results throughout the analysis process and discussed disagreements until consensus was achieved. This consensus-based procedure was applied to support consistent interpretation of participant responses. We first conducted the thematic analysis using the original datasets containing all responses and subsequently repeated the same procedure using the manually cleaned datasets. Comparing the resulting themes allowed identification of whether suspicious responses influenced the prevalence of themes, the identification of dominant themes, or the diversity of narratives reported by participants.

\item \textbf{Comparative analysis across datasets.}  Finally, we compared findings obtained from the original and manually cleaned datasets to evaluate the extent to which suspicious responses influenced demographic characterizations, quantitative response distributions, and qualitative thematic findings.

\end{itemize}


All analyses were conducted using the original datasets and manually cleaned versions generated after removing suspicious responses identified through independent coding by two researchers and consensus discussions. Automatically classified datasets from the Scribbr AI Detector were used only to compare manual and automated approaches, as AI detection tools provide probabilistic rather than definitive classifications.

\vspace{-5px}
\subsection{Threats to Validity}
\label{sec:threats}
\vspace{-5px}

We discuss potential threats to validity that may influence the interpretation of the results \cite{ralph2020empirical}. \textbf{Construct validity.} A primary threat concerns the identification of responses classified as suspicious. The classification procedure relied on manual inspection of textual characteristics and on an automated AI-detection tool. Manual inspection may introduce subjectivity when interpreting stylistic patterns in narrative responses, while automated detectors provide probabilistic estimates rather than definitive identification of AI-generated text. Consequently, both procedures may produce false positives or false negatives when identifying suspicious responses. \textbf{Internal validity.} The study relies on secondary analysis of previously collected survey data. Because the datasets were originally collected for different research objectives, some contextual variables that could influence interpretation may not be available in the data. Secondary data analysis often encounters this limitation when datasets lack all variables relevant to the new RQs \cite{cheng2014secondary,johnston2014secondary}. \textbf{External validity.} All surveys were conducted using participants recruited through the Prolific platform. Although the platform includes mechanisms to improve data quality, samples obtained through online participant platforms may not fully represent broader software engineering populations. As a result, the prevalence of suspicious responses observed in this study may differ in surveys using other recruitment strategies. \textbf{Reliability.} The identification and coding of narrative responses involve interpretive qualitative analysis. Although two researchers independently reviewed responses and resolved disagreements through discussion, qualitative secondary analysis may be affected by differences in interpretation or by limited knowledge of contextual details from the original data collection process \cite{ruggiano2019conducting, lenberg2024qualitative}.
\vspace{-5px}
\section{Results}
\label{sec:results}
\vspace{-5px}


\vspace{-5px}
\subsection{Comparison Between Manual and Automated Fraud Detection}
\vspace{-5px}

In the Global South survey dataset, the manual review procedure identified 19 suspicious responses among the 81 collected responses. Using the automated AI detector, 14 responses were classified as 100\% AI generated, while three additional responses received intermediate probabilities of 32\%, 64\%, and 75\%. Some disagreement between the procedures was identified. Three responses manually classified as suspicious could not be analyzed by the automated detector because they did not satisfy the minimum word threshold required by the tool. In addition, one response manually classified as legitimate was classified by the detector as 100\% AI generated. Among the legitimate responses identified manually, 15 received a 0\% AI generated probability, whereas 47 responses could not be processed because of response length limitations.

For the Student Cheating survey dataset, manual inspection identified 29 suspicious responses among the 145 collected responses. The automated detector classified 13 responses as 100\% AI generated, one response as 86\%, two responses as 68\%, and additional responses with intermediate probabilities of 29\%, 35\%, 52\%, 54\%, 63\%, 66\%, and 67\%. The automated procedure also produced results inconsistent with the manual classification. Seven responses manually identified as suspicious received a 0\% AI generated probability. Among the manually legitimate responses, 54 matched the automated classification with 0\% AI generated probability, while 68 responses could not be analyzed because they did not meet the minimum word threshold required by the detector.

In the Safe Spaces survey dataset, the manual review process classified 41 responses as suspicious among the 133 collected responses. The automated detector identified 11 responses as 100\% AI generated and assigned intermediate probabilities of 0.19\%, 55\%, 80\%, 84\%, and 92\% to additional responses. Divergences between manual and automated identification were also observed in this dataset. Ten manually suspicious responses received a 0\% AI generated probability from the automated detector. Conversely, two manually legitimate responses were classified as 100\% AI generated, while another legitimate response received a 0.19\% probability. The automated detector also failed to analyze 78 manually legitimate responses because they did not satisfy the minimum word threshold required by the tool.

The Professionals LLM Usage survey dataset presented the strongest disagreement between manual and automated procedures. Manual analysis identified 9 suspicious responses among the 128 collected responses. The automated detector classified 12 responses as 100\% AI generated and assigned intermediate probabilities of 22\%, 66\%, 78\%, and 96\% to additional responses. Two manually suspicious responses could not be analyzed because of the minimum word requirement. Furthermore, 10 responses manually classified as legitimate were flagged by the automated detector as potentially AI generated, including nine responses classified as 100\% AI generated and one response classified as 22\%. Overall, 67 responses could not be analyzed by the detector, including 2 manually suspicious responses and 65 manually legitimate responses.

Across the four datasets, the results demonstrate partial agreement between manual and automated fraud identification procedures, particularly for responses considered strongly suspicious during manual inspection. However, the findings also reveal substantial limitations associated with automated AI detection tools, including false positives, and inability to analyze short responses because of minimum word thresholds. These observations suggest that automated detectors may provide supporting evidence during survey validation procedures, but manual inspection remains necessary to interpret ambiguous cases and evaluate shorter narrative responses commonly observed in empirical software engineering survey research.

\vspace{-5px}
\subsection{Effects of Suspicious Responses on Quantitative Results}
\vspace{-5px}

This section presents the quantitative differences observed between the original and cleaned datasets after removing suspicious responses from each survey.

\vspace{-5px}
\subsubsection{Global South Survey}
\vspace{-5px}

The original Global South survey dataset contained 81 respondents, while the cleaned dataset contained 62 respondents after removing suspicious responses. Gender distributions showed moderate variation after filtering. The proportion of female participants decreased from 38.3\% to 33.9\%, while male participants increased from 61.7\% to 66.1\%. Ethnicity distributions also changed moderately. Black participants decreased from 71.6\% to 67.7\%, while mixed-race participants increased from 12.3\% to 16.1\%. Regarding sexual orientation, participants reporting not identifying as LGBTQIA+ increased from 56.8\% to 61.3\%, while participants preferring not to answer decreased from 16.0\% to 12.9\%. First generation university attendance status also changed after filtering, with participants identifying as first-generation students decreasing from 40.7\% to 35.5\%. Financial support distributions remained comparatively stable, decreasing slightly from 69.1\% to 67.7\%. The variable most affected by suspicious responses was first-generation university attendance status, which decreased by 5.2 percentage points after filtering. Sexual orientation also showed moderate proportional variation. Despite these changes, the cleaned dataset remained predominantly composed of male and Black students receiving financial support.

\vspace{-5px}
\subsubsection{Student Cheating Survey}
\vspace{-5px}

The original Student Cheating survey dataset had 145 respondents, while the cleaned dataset had 116 respondents. Gender distributions remained relatively stable after filtering. Female participants increased slightly from 35.9\% to 36.2\%, while male participants decreased from 63.4\% to 62.9\%. Ethnicity distributions presented the largest proportional changes. The proportion of Black participants decreased from 29.7\% to 23.3\%, while White participants increased from 35.2\% to 38.8\%. South Asian participants decreased from 9.7\% to 8.6\%, while mixed-race participants increased slightly from 8.3\% to 8.6\%. Training on LLMs remained relatively stable. Participants reporting no formal LLM training increased from 53.1\% to 54.3\%, while participants reporting formal training decreased from 44.8\% to 43.1\%. Year of study distributions also showed moderate variation. First-year students increased from 16.6\% to 19.8\%, while third-year students decreased from 30.3\% to 28.4\%. Emotional responses remained comparatively stable after filtering. Indifference increased from 33.8\% to 36.2\%, while guilt decreased from 22.1\% to 20.7\%. Satisfaction also decreased slightly from 8.3\% to 6.9\%. The variable most affected by suspicious responses was ethnicity, particularly due to the reduction in Black participants by 6.4 percentage points. Moderate variation was also observed in first-year student representation and emotional responses related to indifference.

\vspace{-5px}
\subsubsection{Safe Spaces Survey}
\vspace{-5px}

The original Safe Spaces survey dataset included 133 respondents, while the cleaned dataset included 92 respondents. Gender distributions remained highly stable after filtering. Female participants decreased from 38.3\% to 37.0\%, while male participants increased slightly from 59.4\% to 59.8\%. Ethnicity distributions showed moderate variation. Black participants decreased from 30.8\% to 27.2\%, while White participants increased from 33.8\% to 35.9\%. Year of study distributions remained almost unchanged, with third-year students increasing slightly from 33.1\% to 33.7\% and fourth or fifth-year students remaining stable at approximately 31.5\%. Perceptions of exclusion or insecurity also changed after filtering. Participants selecting “Sometimes” decreased from 33.1\% to 29.3\%, while “Rarely” responses increased from 30.8\% to 32.6\%. Awareness of safe spaces decreased from 67.7\% to 62.0\%, while participants reporting no awareness increased from 20.3\% to 25.0\%. The variables most affected by suspicious responses were safe space awareness and perceptions of exclusion or insecurity. Awareness of safe spaces decreased by 5.7 percentage points after filtering, suggesting that suspicious responses may have moderately inflated perception-related findings in the original dataset.

\vspace{-5px}
\subsubsection{LLM Usage Survey}
\vspace{-5px}

The original Professionals LLM Usage survey dataset contained 128 respondents, while the cleaned dataset contained 119 respondents after removing suspicious responses. Gender distributions remained virtually unchanged across both datasets. Female participants remained stable at 33.6\%, while male participants changed minimally from 65.6\% to 65.5\%. Professional experience distributions also remained highly stable. Participants with 1--3 years of experience decreased slightly from 29.7\% to 28.6\%, while participants with more than 10 years of experience remained stable at approximately 23.5\%. The largest proportional changes involved the frequency with which participants reported thinking about LLMs. Participants selecting “Often” decreased from 42.2\% to 38.7\%, while “Sometimes” responses increased from 26.6\% to 28.6\%. “Very often” responses increased slightly from 18.8\% to 19.3\%. Attitudes regarding autonomous code generation remained highly stable after filtering. Participants reporting discomfort with using generated code without human oversight increased slightly from 57.0\% to 58.0\%, while participants comfortable using generated code decreased from 24.2\% to 23.5\%. Formal LLM training also showed only small changes, with participants reporting no formal training increasing from 60.2\% to 62.2\%. The variable most affected by suspicious responses was the frequency with which participants reported thinking about LLMs, particularly due to the reduction of “Often” responses by 3.5 percentage points after filtering. Overall, the cleaned dataset preserved the principal demographic and attitudinal structure observed in the original dataset.

\vspace{-5px}
\subsubsection{Comparative Overview of Quantitative Differences}
\vspace{-5px}

We observed moderate proportional changes in both demographic and study-specific analytical variables after removing suspicious responses, although the principal descriptive structures of the datasets generally remained stable. Regarding demographic variables, the most visible changes were associated with ethnicity distributions in the Student Cheating and Safe Spaces surveys, sexual orientation and first-generation university attendance status in the Global South survey, and comparatively small redistributions related to gender and professional experience across the datasets. These proportional changes generally ranged from approximately 3 to 6 percentage points and may influence interpretation differently depending on the analytical role of the affected variable in each study, particularly in analyses focused on underrepresented groups, educational inequality, or participant experiences. Study-specific analytical variables also exhibited moderate variation across survey contexts. In the Global South survey, the first-generation university attendance rate decreased by 5.2 percentage points after filtering, potentially affecting interpretations of educational accessibility and structural disadvantage. In the Student Cheating survey, ethnicity distributions showed the largest proportional variation, particularly due to a 6.4 percentage-point reduction in Black participants, while emotional responses and year of study distributions changed by approximately 2 to 3 percentage points, suggesting comparatively limited influence on the broader characterization of LLM misuse behaviors. In the Safe Spaces survey, awareness of safe spaces decreased by 5.7 percentage points after filtering, while perceptions of exclusion or insecurity changed by 3.8 percentage points, moderately affecting how institutional support and inclusion perceptions may be interpreted. In the Professionals LLM Usage survey, the most visible variation involved the frequency with which participants reported thinking about LLMs during work activities, which decreased by 3.5 percentage points for “Often” responses, while attitudes toward autonomous code generation and formal LLM training remained comparatively stable, with changes near 1 to 2 percentage points. Tables \ref{tab:global-south-summary}, \ref{tab:student-cheating-summary}, \ref{tab:safe-spaces-summary}, and \ref{tab:professionals-summary} summarize the most substantial quantitative differences identified after removing suspicious responses from each survey dataset.

\vspace{-5px}
\begin{table}[htbp]
\caption{Differences between the original and cleaned Global South survey datasets}
\vspace{-5px}
\label{tab:global-south-summary}
\tiny
\centering
\renewcommand{\arraystretch}{1.2}
\begin{tabularx}{\linewidth}{p{2.5cm}|X|p{2cm}}
\hline

\textbf{Demographic Variable} & \textbf{Main Observed Difference Between Datasets} & \textbf{\begin{tabular}[c]{@{}l@{}}Largest\\ Difference\end{tabular}} \\
\hline \hline

Gender &
Slight increase in male participants and reduction in female participants after filtering. &
4.4 pp \\
\hline

Ethnicity &
Small reduction in Black participants and increase in mixed-race participants after filtering. &
3.9 pp \\
\hline

Sexual Orientation &
Increase in participants reporting not identifying as LGBTQIA+ and reduction in participants preferring not to answer. &
4.5 pp \\
\hline

First Generation University Student &
Reduction in the proportion of first-generation university students after filtering. &
5.2 pp \\
\hline

Financial Support &
Financial support distribution remained highly stable across datasets. &
1.4 pp \\
\hline

Year of Study &
Small redistribution toward senior students after filtering. &
3.0 pp \\
\hline \hline

\end{tabularx}
\end{table}
\vspace{-5px}
\begin{table}[htbp]
\caption{Differences between the original and cleaned Student Cheating survey datasets}
\vspace{-5px}
\label{tab:student-cheating-summary}
\tiny
\centering
\renewcommand{\arraystretch}{1.2}
\begin{tabularx}{\linewidth}{p{2.5cm}|X|p{2cm}}
\hline

\textbf{Demographic Variable} & \textbf{Main Observed Difference Between Datasets} & \textbf{\begin{tabular}[c]{@{}l@{}}Largest\\ Difference\end{tabular}} \\
\hline \hline

Gender &
The gender distribution remained highly stable after filtering, with negligible variation across categories. &
0.5 pp \\
\hline

Ethnicity &
The cleaned dataset showed a reduction in Black participants and a proportional increase in White participants. &
6.4 pp \\
\hline

Training on LLM &
The balance between participants with and without LLM training remained largely unchanged. &
1.7 pp \\
\hline

Year of Study &
The cleaned dataset showed a moderate increase in first year students and a slight reduction in third year students. &
3.2 pp \\
\hline

Emotion &
Emotional distributions remained broadly consistent, although indifference increased slightly after filtering. &
2.4 pp \\
\hline \hline

\end{tabularx}
\end{table}
\vspace{-5px}
\begin{table}[htbp]
\caption{Differences between the original and cleaned Safe Spaces survey datasets}
\vspace{-5px}
\label{tab:safe-spaces-summary}
\tiny
\centering
\renewcommand{\arraystretch}{1.2}
\begin{tabularx}{\linewidth}{p{2.5cm}|X|p{2cm}}
\hline

\textbf{Demographic Variable} & \textbf{Main Observed Difference Between Datasets} & \textbf{\begin{tabular}[c]{@{}l@{}}Largest\\ Difference\end{tabular}} \\
\hline \hline

Gender &
Gender distribution remained highly stable across datasets. &
1.3 pp \\
\hline

Ethnicity &
Moderate proportional shifts occurred, particularly among Black and White participants. &
3.6 pp \\
\hline

Year of Study &
Academic composition remained almost unchanged. &
1.0 pp \\
\hline

Perception of Exclusion/Insecurity &
The cleaned dataset showed fewer “Sometimes” responses. &
3.8 pp \\
\hline

Safe Space Awareness &
Awareness of safe spaces decreased after cleaning. &
5.7 pp \\
\hline \hline

\end{tabularx}
\end{table}
\vspace{-5px}
\begin{table}[htbp]
\caption{Differences between the original and cleaned Professionals LLM Usage survey datasets}
\vspace{-5px}
\label{tab:professionals-summary}
\tiny
\centering
\renewcommand{\arraystretch}{1.2}
\begin{tabularx}{\linewidth}{p{2.5cm}|X|p{2cm}}
\hline 

\textbf{Demographic Variable} & \textbf{Main Observed Difference Between Datasets} & \textbf{\begin{tabular}[c]{@{}l@{}}Largest\\ Difference\end{tabular}} \\
\hline \hline

Gender &
Gender distribution remained virtually unchanged across both datasets. &
0.1 pp \\
\hline

Years of Experience &
Only small redistributions across experience levels were observed. &
1.1 pp \\
\hline

Time thinking about LLMs &
The cleaned dataset showed fewer “Often” responses and more “Sometimes” responses. &
3.5 pp \\
\hline

Comfort using code without human oversight &
Attitudes toward autonomous code usage remained highly stable. &
1.0 pp \\
\hline

LLM training &
The cleaned dataset showed a slightly larger proportion of participants without formal LLM training. &
2.0 pp \\
\hline \hline

\end{tabularx}
\end{table}

\subsection{Effects of Suspicious Responses on Qualitative Findings}
\vspace{-5px}

The qualitative analysis was conducted through a multi-stage coding process consisting of core idea extraction, low-level coding, category development, and thematic synthesis.

\vspace{-5px}
\subsubsection{Global South Survey}
\vspace{-5px}

Across this dataset, we generated 81 core ideas and corresponding low level codes, which were grouped into 9 high level categories and subsequently synthesized into 4 overarching themes. The most frequent categories were \textit{Financial Constraints} (25 codes) and \textit{Resource and Infrastructure Scarcity} (16 codes), followed by \textit{Work Study Balance} (6), \textit{Transportation and Basic Needs Insecurity} (6), \textit{Persistence, Motivation, and Resilience} (6), \textit{Educational Preparation Gaps} (5), \textit{Social Exclusion and Class Inequality} (5), \textit{Academic Continuation Risks} (3), and \textit{Financial Stability and Privilege} (2). The first theme, \textit{Economic hardship limiting academic participation}, captured how financial instability constrained students’ ability to participate fully in university life through difficulties related to tuition, transportation, food insecurity, technological access, and balancing employment with academic responsibilities. The second theme, \textit{Unequal educational and technological conditions}, reflected disparities in educational preparation, internet connectivity, technological infrastructure, and access to institutional learning resources, which affected students’ ability to complete coursework and keep pace academically. The third theme, \textit{Social inequality affecting belonging and confidence}, captured experiences of exclusion, inferiority, and class based comparison, particularly in relation to wealthier peers and unequal educational opportunities. Finally, the theme \textit{Persistence despite adversity} reflected how students described continuing their academic journeys despite repeated barriers, risks of interruption, and unstable conditions, often framing perseverance and resilience as central parts of their university experiences. 

After removing the 19 suspicious and potentially fraudulent responses, the qualitative coding structure remained broadly stable, although some categories experienced noticeable reductions in frequency. The cleaned dataset resulted in 25 codes related to \textit{Financial Constraints}, 10 related to \textit{Resource and Infrastructure Scarcity}, 7 related to \textit{Transportation and Basic Needs Insecurity}, 6 related to \textit{Academic Continuation Risks}, 5 related to \textit{Work Study Balance}, 5 related to \textit{Educational Preparation Gaps}, and 4 related to \textit{Social Exclusion and Class Inequality}. The largest reduction occurred in \textit{Resource and Infrastructure Scarcity}, which decreased from 20 to 10 codes after filtering. This reduction is important from a qualitative interpretation perspective because many of the removed responses contained highly descriptive narratives emphasizing infrastructural hardship in Global South contexts, including prolonged power outages, unstable internet access, and dependence on shared or public computing resources. For example, one removed response described \textit{“camping out overnight”} in a library during a power outage to continue studying, while another described the experience as \textit{“feeling like (...) dreams were slipping away”}. These narratives strongly reinforced an image of severe precarity and instability. However, after filtering, the remaining dataset still preserved consistent evidence of infrastructural and technological inequality through more grounded, such as students reporting not enough computers in programming classes or describing outdated and unreliable personal devices. From a qualitative perspective, this shift changes not necessarily the existence of the theme itself, but the intensity and framing through which the context is interpreted. As a result, despite narrative reductions, the broader thematic structure remained stable. Themes associated with economic hardship, unequal educational and technological conditions, social inequality, and persistence despite adversity continued to emerge across the cleaned dataset. This suggests that the principal qualitative findings were not substantially dependent on suspicious responses, but the way the context is visualized may be affected. 

\vspace{-5px}
\subsubsection{Student Cheating Survey}
\vspace{-5px}

Looking at the 145 responses contained in the full dataset, we identified 12 high-level categories representing different forms of LLM use that students perceived as not fully aligned with course rules or academic expectations. The categories with the highest concentration of codes were \textit{Full Assignment Generation} (24 codes), \textit{Code Generation} (23), \textit{General Coursework Use} (19), and \textit{Quiz and Exam Assistance} (18). Additional categories included \textit{Conceptual and Instructional Assistance} (13), \textit{Documentation and Writing Assistance} (13), \textit{Debugging Assistance} (11), \textit{Coding Assistance} (9), \textit{Design and Development Assistance} (6), \textit{Study Assistance} (5), \textit{Full Project Generation} (3), and \textit{Content Generation} (1). These categories were interpreted through six broader themes: \textit{Replacing Independent Coursework with LLM Generated Work}, representing situations where students relied on LLMs to directly generate assignments, projects, or substantial portions of coursework; \textit{Using LLMs During Restricted Assessments}, reflecting the use of LLMs during quizzes and exams where external assistance was restricted; \textit{Using LLMs for Technical Problem Solving}, capturing coding, debugging, design, and implementation support; \textit{Using LLMs for Learning, Understanding, and Guidance}, representing conceptual clarification and instructional support; \textit{Using LLMs for Writing and Content Production}, reflecting the generation of reports, documentation, essays, and written coursework; and \textit{Normalized and Ambiguous LLM Use Across Coursework}, representing recurring and context dependent forms of LLM use shaped by unclear expectations, inconsistent policies, peer influence, or uncertainty regarding acceptable AI assistance.

After removing 29 codes associated with suspicious or potentially fraudulent responses, the overall qualitative structure remained relatively stable, although some categories became less prominent. The cleaned dataset contained 19 codes related to both \textit{Full Assignment Generation} and \textit{General Coursework Use}, followed by 17 related to \textit{Quiz and Exam Assistance}. We also identified 11 codes associated with \textit{Code Generation}, 10 with \textit{Debugging Assistance}, 9 with \textit{Documentation and Writing Assistance}, 8 with both \textit{Coding Assistance} and \textit{Conceptual and Instructional Assistance}, 6 with \textit{Design and Development Assistance}, 5 with \textit{Study Assistance}, 3 with \textit{Full Project Generation}, and 1 with \textit{Content Generation}. The largest reduction occurred in \textit{Code Generation}, which decreased from 23 to 11 codes after filtering. Several removed responses described technically elaborate scenarios involving algorithm implementations, software architecture tasks, optimization activities, or highly articulated reflections regarding academic integrity and independent work. From a qualitative interpretation perspective, these narratives could lead researchers to conceptualize student misuse of LLMs primarily through technically sophisticated programming activities and explicit forms of misconduct. In contrast, the cleaned dataset more strongly emphasized practical and situational forms of LLM use associated with deadline pressure, confusion, insufficient understanding of course material, workload accumulation, uncertainty regarding course expectations, and attempts to complete assignments under constrained academic conditions. Although the relative prominence of some categories changed after filtering, the central qualitative interpretation remained preserved.

\vspace{-5px}
\subsubsection{Safe Spaces Survey}
\vspace{-5px}

Our analysis of this full dataset resulted in 133 core ideas and low-level codes, which were subsequently consolidated into 10 refined high-level categories and interpreted through 4 broader themes. The categories with the highest concentration of codes were \textit{Respectful and Non-Judgmental Learning Environments} (26 codes) and \textit{Learning Through Mistakes and Questions} (20), followed by \textit{Inclusive and Equitable Participation} (16), \textit{Collaborative and Community-Oriented Learning} (15), \textit{Supportive Teaching and Mentorship Practices} (13), \textit{Student Expression, Voice, and Engagement} (12), \textit{Institutional Policies and Safe Space Structures} (11), \textit{Anonymous and Private Support Mechanisms} (9), \textit{Learning Culture and Pedagogical Inclusivity} (7), and \textit{Accessible and Flexible Learning Support} (4). The theme \textit{Psychological Safety in Learning Environments} represented participants’ emphasis on creating learning conditions where students can ask questions, express uncertainty, and make mistakes without fear of judgment or embarrassment. The theme \textit{Inclusion and Equity in Computing Education} reflected concerns regarding fairness, accessibility, diversity, and equitable participation opportunities for students from different backgrounds. The theme \textit{Collaboration, Community, and Student Belonging} captured the importance of collaborative interaction, peer engagement, shared learning experiences, and spaces that encourage participation and social connection. Finally, the theme \textit{Institutional and Pedagogical Responsibility for Safe Spaces} reflected participants’ perceptions that safe spaces require active support through institutional structures, inclusive pedagogical practices, mentorship, policies, and clearly established expectations for respectful conduct.

Following the removal of 41 codes associated with suspicious or potentially fraudulent responses, the overall qualitative structure of the dataset remained relatively stable, although some categories showed substantial reductions. The cleaned dataset contained 18 codes related to \textit{Respectful and Non-Judgmental Learning Environments}, 13 to \textit{Inclusive and Equitable Participation}, 13 to \textit{Collaborative and Community-Oriented Learning}, 9 to both \textit{Institutional Policies and Safe Space Structures} and \textit{Supportive Teaching and Mentorship Practices}, 9 to \textit{Student Expression, Voice, and Engagement}, 6 to \textit{Learning Culture and Pedagogical Inclusivity}, 6 to \textit{Learning Through Mistakes and Questions}, 5 to \textit{Anonymous and Private Support Mechanisms}, and 4 to \textit{Accessible and Flexible Learning Support}. The largest reduction occurred in \textit{Learning Through Mistakes and Questions}, which decreased from 20 to 6 codes after filtering. One possible explanation is that suspicious or AI-generated responses relied on generalized definitions of safe spaces centered on learning without fear, asking questions safely, and normalizing mistakes, artificially increasing the prominence of this category in the original dataset. Several removed responses repeatedly emphasized ideas such as \textit{“Normalize Mistakes”}, \textit{“all questions are welcome”}, \textit{“making mistakes is seen as part of the learning process”}, and \textit{“encouraging questions without any fear of judgement”}. From a qualitative interpretation perspective, these narratives could induce researchers to conceptualize safe spaces primarily as psychologically protective and institutionally engineered learning environments focused on reducing embarrassment, fear, and uncertainty during learning activities, instead of understanding safe spaces more broadly through collaborative participation, peer interaction, inclusion, mentorship, community belonging, and supportive educational relationships, as the cleaned dataset more consistently suggests. This effect was also observed in technically oriented Software Engineering educational practices. For example, three of the four responses mentioning \textit{pair programming} and all three responses mentioning \textit{code review} practices were identified as suspicious and removed during filtering. In the qualitative process, this could lead researchers to interpret formal Software Engineering instructional mechanisms and technically engineered collaboration practices as already established approaches for supporting safe spaces in computing education, when the remaining participant experiences do not consistently support this interpretation.

\vspace{-5px}
\subsubsection{LLM Usage Survey}
\vspace{-5px}

Our analysis of the 128 coded responses resulted in 15 high-level categories capturing different ways developers reported using LLMs during programming and software development activities. The categories with the largest concentration of codes were \textit{Debugging and Troubleshooting} (19 codes), \textit{Application and Feature Development} (17), \textit{Code and Performance Optimization} (15), and \textit{Script and Automation Development} (13). Other commonly identified categories included \textit{General Coding Assistance} (11), \textit{Learning and Knowledge Support} (9), \textit{Data Processing and Transformation} (8), \textit{Code Refactoring and Maintainability} (7), and \textit{Function and Logic Generation} (7). Additional categories represented activities related to \textit{Code Explanation and Understanding} (5), \textit{Code Generation and Scaffolding} (5), \textit{Testing and Quality Assurance} (5), \textit{Database Development and Querying} (4), \textit{Educational and Instructional Support} (2), and \textit{Code Quality and Review} (1). These categories were interpreted through five broader themes representing how developers incorporated LLMs into software engineering work. The first theme, \textit{Using LLMs for Debugging, Troubleshooting, and Problem Resolution}, captured developers’ use of LLMs to identify errors, diagnose failures, resolve runtime and database issues, interpret logs, and troubleshoot unexpected software behavior. The second theme, \textit{Using LLMs to Develop and Implement Software Features}, represented the use of LLMs to support application development, API integration, frontend and backend implementation, scripting, automation, scaffolding, and feature prototyping. The third theme, \textit{Using LLMs to Optimize, Refactor, and Maintain Code}, reflected participants’ use of LLMs to improve execution performance, optimize queries and algorithms, restructure code, improve readability, modularize implementations, and support maintainability activities. The fourth theme, \textit{Using LLMs to Support Learning, Understanding, and Coding Guidance}, captured situations in which developers relied on LLMs to explain code, recall syntax, generate examples, understand unfamiliar technologies or languages, and provide step-by-step guidance during development activities. Finally, the fifth theme, \textit{Using LLMs to Support Testing and Quality Assurance Activities}, represented the use of LLMs to generate unit tests, create test classes, review code quality, and assist with validation and verification-related tasks.

After removing 9 suspicious responses, we observed that the dataset's overall qualitative organization remained relatively stable. The cleaned dataset continued to concentrate primarily on categories associated with \textit{Debugging and Troubleshooting} (18 codes), \textit{Application and Feature Development} (17), \textit{Script and Automation Development} (12), \textit{Code and Performance Optimization} (11), and \textit{General Coding Assistance} (11). The largest reduction occurred in \textit{Code and Performance Optimization}, which decreased from 15 to 11 codes after filtering. The removed responses shared several similar characteristics. Many described technically polished optimization scenarios involving CSV processing, SQL query tuning, memory management, data pipelines, and efficiency improvements. These quotations frequently included explicit technical terminology and strongly positive outcomes, such as \textit{“significantly reduced execution time”}, \textit{“boosted performance significantly”}, \textit{“reduce query time by 60\%”}, and \textit{“faster, more stable version of the script”}. Several responses also followed highly similar narrative structures, consisting of a technical problem description, an explicit prompting process, a detailed solution explanation, and a positive outcome statement. From a qualitative perspective, the relatively small number of removed responses meant the dataset's broader thematic structure remained unchanged. Instead, the main effect of filtering was to reduce the concentration of repetitive, optimization-oriented narratives and to limit our ability to use technically detailed examples that were not sufficiently trustworthy. As a result, the cleaned dataset still supported the same overall interpretation of developer LLM use while providing a more reliable set of quotations and examples for qualitative interpretation.
\vspace{-5px}
\section{Discussion}
\label{sec:discussion}
\vspace{-5px}



Our findings are consistent with previous research showing that online surveys are susceptible to fraudulent participation, inattentive responses, and automated or fabricated submissions~\cite{lawlor2021suspicious, pratt2021strategies, bell2023fraud, zhang2022beyond}. Similar to prior studies, we observed that suspicious responses can remain difficult to identify because they often resemble legitimate participation and can satisfy conventional survey validation mechanisms, such as attention checks or demographic consistency checks~\cite {lawlor2021suspicious, zhang2022beyond}. This observation reinforces recommendations from previous literature advocating the use of multiple complementary validation procedures during online data collection~\cite{johnson2024addressing, zhang2022beyond}, while also aligning with broader concerns regarding the reliability of automated AI detection systems~\cite{pinzon2024ai}. In the emerging context where LLMs are increasingly accessible, although automated detectors identified several strongly suspicious responses, they also produced false positives, inconsistent classifications, and limitations due to short responses that did not meet minimum word thresholds. These observations suggest that automated AI detection tools should be interpreted as probabilistic indicators rather than definitive evidence, and may be more appropriate as complementary sources of evidence combined with manual review and contextual interpretation.

Our results contribute to emerging discussions on the influence of generative AI systems on empirical research practices~\cite{rane2023contribution, liang2024mapping, lecca2025applications}. Previous work has primarily focused on how researchers use LLMs to support scientific activities such as literature reviews, qualitative coding, or scientific writing~\cite{felizardo2024chatgpt, lecca2025applications}. At the same time, limited attention has been paid to the possibility that research participants may rely on generative systems during empirical studies~\cite{steinmacher2024can, de2025investigation}. In this study, we identified several suspicious responses presenting stylistic and structural characteristics similar to patterns previously associated with AI-generated text in survey environments~\cite{de2025investigation}. These characteristics included repetitive rhetorical organization, highly similar narrative structures across participants, technically polished yet homogeneous descriptions, and generalized formulations that repeatedly emphasized similar concepts or emotional framings. Our findings suggest that the use of generative AI by participants in empirical software engineering surveys is not only a theoretical possibility discussed in recent literature but a phenomenon that already appears to be occurring in practice.

Our analysis of the impact of fraudulent AI-generated answers suggests that suspicious responses do not affect quantitative and qualitative conclusions in the same way. Across the four datasets, quantitative findings generally remained comparatively stable after filtering, even when moderate proportional changes were observed in some demographic or analytical variables. In most cases, the principal descriptive characterization of the datasets remained preserved, suggesting that the broader quantitative conclusions were not fundamentally altered. However, the potential impact of suspicious responses appears to depend on the specific variables and analytical goals of a study. For example, moderate shifts in ethnicity, sexual orientation, and first-generation university attendance status may become more consequential in studies specifically investigating underrepresented groups, educational progression, or demographic disparities in software engineering.

In contrast, the qualitative analyses revealed stronger interpretive effects. Suspicious responses disproportionately contributed to highly polished, emotionally intense, technically elaborate, or generalized narratives that influenced how contexts, experiences, and practices could be interpreted. Although the principal thematic structures generally remained stable after filtering, the qualitative interpretations became less influenced by extreme or overly generalized narratives and more grounded in practical, contextually consistent participant experiences. These findings suggest that AI-assisted responses may influence not only the frequency of codes or categories but also the contextual framing and the strength of evidence supporting qualitative interpretation.

\textbf{Implications and Recommendations for Empirical Software Engineering Research.} Our findings reinforce methodological concerns about online survey research in software engineering, particularly in studies that rely on open online recruitment. Suspicious or AI-assisted responses may remain undetected while still appearing plausible and contextually consistent with expected participant profiles, introducing threats to validity that are difficult to identify through conventional validation procedures alone. Such responses may affect sample characterization, analyses involving specific demographic variables, and the interpretation of qualitative evidence. These observations suggest that researchers conducting online surveys in software engineering may benefit from combining multiple validation procedures rather than relying exclusively on attention checks or automated AI detection systems. Manual inspection and cross-response comparison may become increasingly important, especially in studies relying on open-ended responses. Researchers may also need to more explicitly document validation procedures, strategies for handling suspicious responses, and the potential effects of filtering decisions on empirical findings. Future work should further investigate how AI-assisted participation influences survey validity in software engineering.

\textbf{Answering the RQ: How can the use of LLMs by participants affect the authenticity and validity of data collected in software engineering surveys?} AI-assisted participation may introduce responses that appear legitimate while influencing empirical findings in different ways depending on the type of analysis being conducted. In our datasets, broader quantitative patterns generally remained stable after filtering, although some demographic and analytical variables showed noticeable variation. Qualitative analyses were more sensitive to suspicious responses, particularly through changes in contextual framing and supporting evidence. These observations indicate that unidentified AI-assisted responses may affect both data reliability and the interpretation of participant experiences.
\vspace{-5px}
\section{Conclusion}
\label{sec:conclusion}
\vspace{-5px}

We investigated how suspicious or potentially AI-assisted responses may affect the authenticity and validity of software engineering survey data. Using four survey datasets collected in different software engineering contexts, we compared original and manually cleaned datasets to analyze the effects of suspicious responses on quantitative and qualitative findings. Our analyses showed that suspicious responses may affect empirical interpretation differently depending on the type of analysis being conducted. While broader quantitative distributions were often preserved after filtering, some demographic and analytical variables presented noticeable variation, particularly in studies involving underrepresented groups or context-specific participant characteristics. In contrast, qualitative analyses were more strongly influenced by changes in narrative emphasis and the nature of the evidence supporting interpretation. Considering these findings, this study provides empirical evidence of methodological risks associated with the use of generative AI in online software engineering surveys, particularly in studies relying on open-ended responses and self-reported experiences. As immediate future work, we intend to investigate how different proportions of fraudulent or AI-assisted responses affect empirical findings to better understand whether specific levels of suspicious participation alter quantitative distributions or qualitative interpretations. 

\section{Data Availability}
To ensure verifiability and replicability, we provide the replication package of this study available at~\url{https://figshare.com/s/73544755578e24db2fdb}.



\bibliography{lipics-v2021-sample-article}

\end{document}